\documentclass[aps,preprint,amsmath,amssymb,showpacs,amsfonts]{revtex4}
\usepackage{epsfig}
\usepackage{graphicx}
\usepackage{dcolumn}
\usepackage{bm}

\newcommand{\beq}{\begin{equation}}
\newcommand{\eeq}{\end{equation}}
\newcommand{\bea}{\begin{eqnarray}}
\newcommand{\eea}{\end{eqnarray}}
\def\half{{\scriptstyle{\frac{1}{2}}}}

\begin{document}

\title{Gravity and a Geometrization of Turbulence: An Intriguing Correspondence}

\author{Christopher Eling$^1$}
\author{Itzhak Fouxon$^2$}
\author{Yaron Oz$^2$}

\affiliation{$^1$ SISSA, via Bonomea 265, Trieste 34136, Italy;
INFN Sezione di Trieste, via Valerio 2, Trieste 34127, Italy}

\affiliation{$^2$ Raymond and Beverly Sackler School of
Physics and Astronomy, Tel-Aviv University, Tel-Aviv 69978, Israel}

\date{\today}

\begin{abstract}
The  dynamics of fluids is a long standing challenge that remained as an unsolved problem for centuries.
Understanding its main features, chaos and turbulence, is likely to provide an understanding of the principles and
non-linear dynamics of a large class of  systems far from equilibrium.
We consider a conceptually new viewpoint to study these features using black hole dynamics.
Since the gravitational field is characterized by a curved geometry,
the gravity variables provide a geometrical framework for studying  the dynamics of fluids:
A {\it geometrization} of turbulence.

\end{abstract}
%
\maketitle

\section{Introduction}

Turbulence is often considered as the last major unsolved problem of classical physics \cite{Frisch}. Despite centuries of research,
an analytical description and understanding of fluid flows in the non-linear regime are still lacking. This is while such flows are rather generic in nature:
they arise ubiquitously in everyday situations and represent the absolute majority of fluid motions. These flows pervade our world at all scales:
in the air in the room we sit in, in aircraft motions, in atmospheric phenomena and weather, in astrophysics and cosmology and so on.
Insights to turbulence hold the key to a
tremendous number of technological problems. Yet, our knowledge of the phenomenon is quite modest.

The problems posed by the Navier-Stokes (NS) equations, which provide a mathematical formulation of the flow evolution are  formidable.
An
understanding of the solutions to these equations was chosen by the Clay Mathematics Institute as one of
the seven millennium problems, and one
of the two problems originating from physics.
The main source of difficulty when analyzing the NS equations is the strong non-linear interaction and the absence of a small
expansion parameter.
The interaction holds at a wide range of scales of the flow and
leads to its very complex organization.

The interaction is classical, and in one of the basic formulations is described by a non-linear system of partial differential equations (PDE) that govern the incompressible (solenoidal) velocity field of the flow.
This formulation describes fluid flows with a small Mach number, i.e. where the characteristic velocity is much smaller than the speed of sound.
An appropriate limiting
procedure, applied to the complete set of fluid dynamical equations, allows one to perform
a reduction in the number of independent fields from five (three components of the velocity field and two independent thermodynamical fields characterizing the change of the conditions of local thermal equilibrium throughout the fluid) to four - velocity and pressure. Moreover, the
pressure appears essentially as a constraint ensuring the incompressibility of the velocity field. By knowing the velocity field at a given time, one can determine the pressure at this time from a Poisson equation with the source determined by the velocity. As a result, the NS equations to a large extent describe an evolution of the flow field that provides the distribution of the velocities of fluid particles throughout space. This evolution is simply the second law of Newton, "acceleration equals force", written for every fluid particle, with the field of the pressure force
expressed via the field of velocity. Here, one takes into account the fact that in the limit of small Mach numbers the fluid density is constant. While the physics
that dictates the NS equations is clear, its space of solutions is not understood. The inertial forces make the
equations non-linear, and an analytical analysis of the NS equations is currently possible only in a few special cases, where these forces
effectively vanish and linear dynamics applies.

In the non-linear regime the
velocity field exhibits a highly complex spatial and temporal structure: spatial and temporal snapshots of the field reveal characteristic bursts accompanied by large relatively calm regions, a property referred to as intermittency. The turbulent velocity field appears to be a random process.
For example, single-point measurements of the velocity field give an extremely complex signal that looks random. Thus, a single realization of a solution to the NS equations is considered to be unpredictable.
The attention of the theorists has shifted to the study of average properties of an ensemble of solutions.

For instance, instead of making attempts to understand the whole picture of a single-time velocity field, one considers the statistics of velocity differences between points separated by a fixed distance. The statistics is defined by averaging over the space.
Note, that even when turbulence is excited by a time-independent source, still at large non-linearity the solution
depends non-trivially on time. In contrast, the averages are time-independent and describe a certain steady-state statistics.
The averages exhibit some resemblance  to the physics of systems at equilibrium, with the exact velocity field and the averages being analogous
to the microscopic dynamics and the ``macroscopic properties", respectively.

There are strong indications that statistical average properties are analytic and can be predicted (\cite{Frisch} and references therein). Moreover, it seems plausible that the averages exhibit some degree of universality with certain aspects shared by all turbulent flows, independently
of the details of the flow excitations. The theoretical study of averages is performed by considering statistical ensembles of velocity fields generated by the NS equations. While some significant insights into turbulence have been obtained via the statistical approach \cite{Frisch,FGV,FalkovichSreenivasan}, most of the difficulties remained.
One is required to consider the statistics of a non-linear random field, which has the degree of complexity of a strongly coupled non-linear quantum field theory. It is obvious that the study of the solutions of the NS equations and their statistics requires new ideas and new approaches.

In this brief review we will present recent progress in the development of a new geometrical approach to the problem of turbulence.
While the idea of using geometrical methods to study the NS equations is an old one \cite{A,AK}, the present approach originates from a very different angle,
that is via a holographic relation to gravity.
The classical gravitational force is described
by a system of non-linear PDE - Einstein equations of general relativity \cite{MTW}.
A hint for a possible link between the two non-linear systems comes from that fact that the NS equations that govern the fluid velocity field describe also volume-preserving motions (diffeomorphisms) of space.
Such a motion induces an evolution of the metric that describes the distances between the fluid particles. On the other hand, the theory of general relativity without sources also allows a formulation as an evolution of the spatial metric, though the dynamics is of a more general nature \cite{MTW,ADM}. Thus, it
it is possible that a certain subclass of solutions of Einstein equations without sources is equivalent to the motions described by the NS equations. The relation cannot be exact, for instance because of the difference in the symmetries of the two theories: Galilean invariance compared to local Lorentz invariance. However, as we will show it can hold in a certain approximation.
Thus, one may say that the forces acting inside the fluid can be described geometrically, analogously to the way general relativity describes the force of gravity.

The review is organized as follows. In the next section we will describe the general framework of hydrodynamics and the phenomenon
of turbulence in a setting appropriate for the discussion of the correspondence to gravity. We will present known theoretical and experimental
results on turbulence
including some very recent ones. In section III we will discuss the holographic principle and the connection between black hole horizon dynamics
and hydrodynamics.
In the last section we will discuss the implications of the relation between fluid dynamics and black holes with an
outlook to the future.

\section{Hydrodynamics and Turbulence}

\subsection{Hydrodynamics as a universal description of many-body systems}

Hydrodynamics provides a universal description of the evolution of a class of many-body systems at large temporal and spatial scales \cite{Forster,Landau10}. The universality
of this description has roots similar to those of thermodynamics. The main paradigm of hydrodynamics starts with a system of exact microscopic
conservation laws:
\begin{eqnarray}&&
\partial_t q^a+\nabla\cdot j^a=0 \ ,
\end{eqnarray}
where $q^a, a=1,...,N$ are charges and $ j^a$ are currents. These conservation  laws, satisfied either classically or by the quantum-mechanical averages,
play a special role in the evolution of the system,  since they provide degrees
of freedom that evolve at an arbitrarily large time-scale.

Indeed, a spatial Fourier transform of this equation yields that the rate of change of the Fourier components of $q^a$ is proportional to the wave-number.
With an appropriate assumption of analyticity this implies
that the small wave-number components of $q^a$ are slow variables. Thus, at long time-scales  the large-scale components of the charges are part of the
effective degrees of freedom of the dynamics. In a normal fluid, the locally conserved charges of energy-momentum and the number of particles provide
the complete list of slow degrees of freedom. The system relaxes to equilibrium by first achieving a state of local thermal equilibrium (LTE) within a time-scale $\tau$ (say for a gas $\tau$ is the mean free time, i. e. the mean free path divided by the typical molecular velocity which is the
thermal velocity. For air mean free path is of order of tenth of micron, while thermal velocity is of order of hundreds of meters per second). At $t\gg \tau$ the fields $q^a$ vary on a spatial scale $L_s$ that is much larger than the correlation length $l_{cor}$ (which for gas is just the mean free path), and the
dimensionless Knudsen number
\begin{equation}
Kn\equiv l_{cor}/L_s \ ,
\end{equation}
is much smaller than one. Generally $L_s$ is the minimal scale at which the parameters of LTE vary in space.
At $Kn\ll 1$ one can use the hydrodynamics equations that describe the evolution of $q^a$.

The hydrodynamics equations
form a closed set since the system's degrees of freedom are $q^a$ and the currents become local functionals of $q^a$. The
expressions for the currents $j^a$ in terms of $q^a$ are called constitutive relations. They take the form of a series in
the small parameter $Kn\ll 1$,
\begin{eqnarray}&&
j_i^a=F^a_i(\{q\})+\sum_{jb}G^a_{i, jb}(\{q\})\nabla_j q^b+\ldots \ , \label{exp}
\end{eqnarray}
which is effectively an expansion in gradients, where dots refer to higher order terms involving either higher order derivatives or powers larger than one
of first order derivatives.

In general, the zeroth order, reactive, $F^a_i(\{q\})$ term leads to a conservative dynamics. In a general frame of reference this term is present even in the global thermal equilibrium state when the system is uniform. The momentum current would involve momentum density times the velocity of the uniform motion of the system plus a term related to the work done by pressure. As such this term does not lead to dissipation. To account for dissipative effects, that lead to homogenization if the system is originally non-uniform, one has to consider the first order term.
An example of such a term is an internal friction force that tends to  homogenize the velocity of nearby regions in the fluid.
It is proportional to velocity gradient as measure of velocity non-uniformity.
Often, as we will assume below, it is sufficient to keep only the first two terms in the above series.
The gradients being small correspond to $q^a$ having, effectively, only low wave-number Fourier components that evolve slowly.
 The resulting evolution equations
\begin{eqnarray}&&
\partial_t q^a+\nabla\cdot j^a=f^a \ ,\label{hyd1}
\end{eqnarray}
where $j^a$ are expressed in terms of $q^a$, and we allowed for the external (random) source fields $f^a$, are the equations of hydrodynamics. The currents $j^a$ characterize the internal state of the fluid, while the force $f^a$ characterizes the external conditions.

\subsection{Turbulence}

In general there are five hydrodynamics equations, that consist of four equations for the energy-momentum densities, and one equation for
the density of particles. An important reduction can be done for flows with low Mach number, i.e. where the ratio of the characteristic
velocity difference $V$ of the flow and the speed of sound is small. Since the speed of sound is always smaller than the speed of light, flows with low
Mach number are non-relativistic. One can reduce then the complete system of hydrodynamics equations to
the incompressible
Navier-Stokes (NS) equations

\begin{equation}
\partial_t v_i + v_j\partial_j v_i=-\partial_i P+\nu\partial_{jj} v_i + f_i,\ \ \  \partial_i v_i=0 \ , \label{NS}
\end{equation}
where $v_i(x, t), i=1,...,d,$ ($d \geq 2$) is the velocity vector field,  $P(x,t)$ is the fluid pressure
divided by the density, $\nu$ is the (kinematic) viscosity and
$f_i(x,t)$ are the components of an externally applied (random) force. The above equations are obtained in the "double limit" where first one
takes the limit of small Knudsen number and then of small Mach number. Viscosity describes internal friction between nearby layers of the fluid
in motion with respect to each other.

The NS equations provide a fundamental formulation
of the nonlinear dynamics of fluids.
With a divergence free
force, one can express the pressure via the velocity by taking the
divergence of the NS equation.
The equations can be studied
mathematically in any space dimension $d$, with two and three
space dimensions having an experimental realization.

The flows are characterized by a
dimensionless parameter - the Reynolds number, measuring the ratio of the non-linear to viscous terms in the equation,
\begin{equation}
{\cal R} = \frac{L V}{\nu} \ ,
\end{equation}
where $L$ is the characteristic length scale of velocity difference (in typical everyday conditions $L$ is of the order of meters and $V$ of meters per second). Note, that in general $L$ is not the same as $L_s$ from the previous section. For instance,
for turbulence, velocity profile has typical velocity difference $V$ occurring at the (correlation) scale of the external forcing $L$, while
$L_s$ is determined by the viscosity.
For an incompressible flow with a given geometrical shape of the boundary,
the Reynolds number is the only control parameter (similarity principle).
Experimental and numerical analysis data show that for ${\cal R} \ll 1$, the flows are regular (laminar).
For a Reynolds number in the range between $1$ and $100$ the flow exhibits a complicated (chaotic) structure, while for
 ${\cal R} \gg 100$ the flow is highly irregular (turbulent) with a complex spatio-temporal pattern formed by the turbulent velocity
field.
Most flows in nature are turbulent.
This is simple to see by noting that the viscosity of water is $\nu \simeq 10^{-6} \frac{m^2}{sec}$, while that of air is
 $\nu \simeq 1.5 \times 10^{-5} \frac{m^2}{sec}$.

As explained in the introduction, when studying turbulence one shifts the focus from individual solutions of the NS equations to the
statistics of the solutions. A formal setup can be realized by considering $f_i$ in Eq.~(\ref{NS}) as a stationary random process, say Gaussian,
and taking
the limit of large correlation length $L$ of the force.
The NS equations can be viewed as a non-linear map from the force to the velocity field, with statistics of the velocity being determined by the
statistics of the force.
Constructing the stationary statistics of the solutions for a given statistics of the force
is a difficult task. Essentially, the only exact non-trivial relations follow from the general statistical
law shown for the equation of the type of Eq.~(\ref{hyd1}) in \cite{FFO}.
It was demonstrated there that in the range of distance scales $l \ll r \ll L$ (the inertial range), where $l$ is determined by the viscosity, one has a linear
scaling law for a correlation function, (we define $F^a_i(\bm r)\equiv F^a_i(\{q(\bm r)\})$
\begin{eqnarray}&&
\langle q^a(0) F^a_i(\bm r)\rangle =\epsilon_f r_i \ , \label{scaling}
\end{eqnarray}
where angular brackets stand for averaging over the statistics of the force, assumed to be isotropic, and $\epsilon_f$ is a constant characterizing
the forcing.

Equation (\ref{scaling}) can be used for any charge. When applied to the momentum density $v_i$, which in the limit of small Mach
is just the velocity since
the number density is a constant, one reproduces a relation equivalent to the celebrated Kolmogorov law \cite{Kol},
\begin{equation}
S_3(r)=-\frac{4}{5}\epsilon r \ . \label{Kolm}
\end{equation}
Here $\epsilon$ is the mean rate of energy dissipation per unit volume due to viscosity, and we introduced the longitudinal structure functions of order $n$, (${\bf r} \equiv {\bf x} - {\bf y}$)
\begin{equation}
S_n(r) \equiv \left\langle \left(({\bf v}({\bf x})-{\bf v}({\bf y}))\cdot \frac{{\bf r}}{r}\right)^n\right\rangle \ .
\end{equation}

Until recently, the Kolmogorov "four-fifths law" expressed by Eq.~(\ref{Kolm}) has been the only non-trivial known result on the statistics.
However, Eq.~(\ref{scaling}) can also be applied to another charge, the energy density $v^2/2$ which current contains $pv_i$ term. This leads
to a new relation, a second exact scaling law for turbulence,
\begin{equation}
\langle v_i(r)p(r)v^2(0) \rangle = {\tilde \epsilon} r_i \ , \label{new}
\end{equation}
where ${\tilde \epsilon}$ is a constant determined by averages like $\langle v^2(\bm f\cdot \bm v)\rangle$ (see \cite{FFO} for details). Relation (\ref{Kolm}) is often interpreted in terms of the energy cascade picture, which
says that the energy injected into the fluid by the forcing at a scale $L$ is transmitted to velocity fluctuations at smaller scales
by a non-linear instability process. This leads to a continuous fracturing of the velocity fluctuations into smaller scales
until a viscous scale $l$ is reached, where the energy dissipates due to friction. However, the Kolmogorov law
(\ref{Kolm}) is the small Mach number limit of a general statistical law holding for compressible turbulence with finite Mach number.
The latter law, derived  in \cite{FFO}, follows from the stationarity condition
of the pair correlation function of momentum, and has no obvious relation to the energy balance.
This puts a question mark on the relevance of the cascade picture.

For completeness, we briefly sketch the argument of
\cite{FFO} that applies to turbulence in a barotropic fluid, where the
pressure  $p(\rho)$  is a function of the fluid density $\rho$
only. The hydrodynamic equations have the form
\begin{eqnarray}&&
\partial_t\rho+\nabla\cdot(\rho\bm v)=0,\ \ \
\partial_t(\rho v_i)+\partial_j (\rho v_iv_j+p\delta_{ij}) 
=-\partial_j \left[G^i_{j, kb}(\{\rho\})\nabla_k
\rho^b\right]+f^i, \label{a2}
\end{eqnarray}
where the source often has the form $f^i=\rho \nabla_i \Phi$
that corresponds to an external potential $\Phi$, and the
exact form of $G^i_{j, kb}$ is not important below. The above equations reduce to Eqs.~(\ref{NS}) in the limit of
small Mach number.
A Kolmogorov-type relation for the above system remained elusive for many decades. It was finally obtained on the basis
of the general relation (\ref{scaling}) which application to $q^a=\rho \bm v^a$ gives
\begin{eqnarray}&&
\sum_j\left\langle \rho(0)v_j(0)\left[\rho(\bm r)v_j(\bm r)v_i(\bm
r)+p(\bm r)\delta_{ij} \right]\right\rangle=\frac{\epsilon r_i}{d}
\label{Kolmcompr}
\end{eqnarray}
where $\epsilon$ is defined in this case as $\langle \rho(0)\bm
v(0)\cdot \bm f(0)\rangle=\epsilon$  (we summed over $j$ to get a
more symmetric result) \cite{FFO}. The above relation was probably not derived before
because it requires the consideration of the steady-state condition for the
fourth-order correlation function $\langle \rho(0, t)v_j(0,
t)\rho(\bm r, t)v_j(\bm r, t)\rangle$, while usually in trying to
find Kolmogorov type relations one considers steady state
conditions for the second moment. In the limit of small Mach number
one has $\rho=1$ and $\nabla\cdot\bm v=0$. Then the pressure term is zero and (\ref{Kolmcompr}) is reduced to
\begin{eqnarray} &&
\langle v_{i}(\bm r)v_{j}(\bm r)v_{j}(0)\rangle=\epsilon
r_i/d\,,\label{d1}
\end{eqnarray}
which is equivalent to the usual Kolmogorov relation for incompressible turbulence. Hence  (\ref{Kolmcompr}) is
indeed a general form of the Kolmogorov relation for an arbitrary
Mach number. As we see, from a general viewpoint, the relation
follows from the stationarity of the pair correlation function of
the momentum density rather than from the energy spectral density.
In particular, $\epsilon$ in (\ref{Kolmcompr}) is the input rate of the
squared momentum   and not that of the energy, which coincide (up
to the factor $1/2$) only in the incompressible case.

The Kolmogorov law is not invariant under the substitution $\bm v\to - \bm v$, and thus it breaks the time-reversal symmetry.
This breaking occurs in the interval of scales where viscous dissipation is irrelevant, reflecting that the steady state assumption
used by the relation, anticipates  that at the very smallest scales there is an energy sink. The relation
has been confirmed by experiment. The unique feature of this
law and the law described by Eq.~(\ref{new}) is universality. Both laws hold in the inertial range independently of the statistics
of the forcing.
The only place where the forcing enters
is in the two multiplicative constants measuring the amplitude of the fluctuations.

It is natural to inquire, whether the
statistics in the inertial range is universal. Indeed, within the cascade picture, velocity fluctuations in this range are formed by many
independent instabilities that fracture the velocity field and thus universality could be expected. In particular,
one may try to conjecture that the statistics is self-similar in the inertial range, leading to $S_n(r) \propto r^{n/3}$.
There is experimental and numerical evidence that in the inertial range, the flows exhibit a universal behavior (\cite{Frisch} and references therein)
\begin{equation}
S_n(r) \sim r^{\xi_n} \ ,
\label{expon}
\end{equation}
where $\xi_n$ is a universal function independent of the statistics of the forcing.
However, in three space dimensions $\xi_n$ is not a linear function of $n$ and self-similarity is broken (see Figure 1).
The calculation of the anomalous exponents $\xi_n-n/3$ is a major open problem. Let us note that in two space dimensions, where the scaling range holds at $r\gg L$ and not $r\ll L$, the exponents $\xi_n$ do seem to follow the $n/3$ law.

In order to explain the deviation of $\xi_n$ from linearity,  several phenomenological models have been
proposed \cite{Frisch}. 
\begin{figure}[htb]
\begin{center}
\epsfig{file=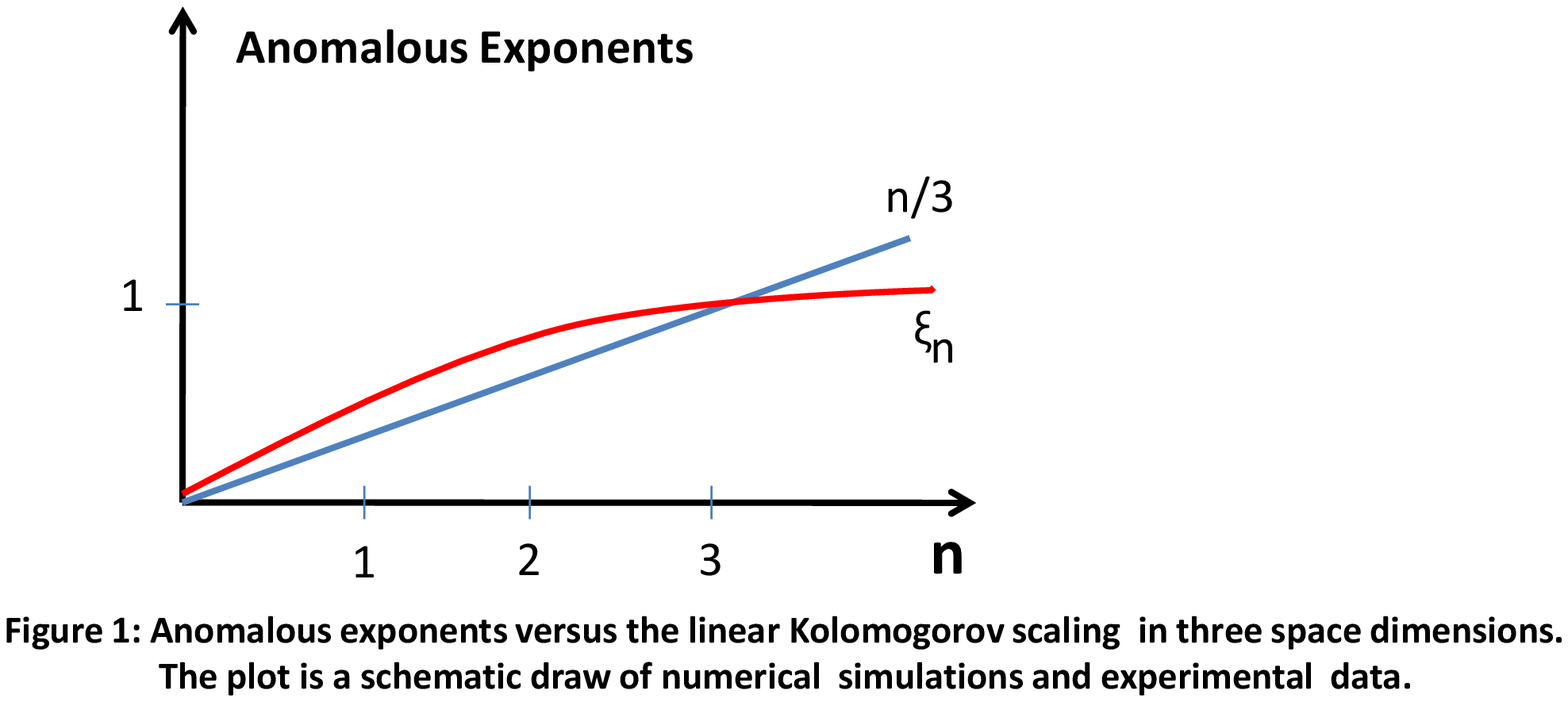,height=11cm}
\end{center}
\label{factor}
\vskip -6cm
\end{figure}
The most successful model is the so-called multi-fractal model of turbulence.
In its original formulation one postulates that the space occupied by the flow can be divided
into fractal subsets labeled by a continuous index $\alpha$, that takes values in an interval.
It is assumed that within each subset self-similarity, which is generally broken in the
whole space, does hold.
Space averages such as the structure functions $S_n(r)$
are given in terms of sums over the self-similar subsets.
The velocity moments of different order are dominated
by different $\alpha$'s resulting in non-linear $\xi_n$.
The non-linearity of $\xi_n$ is determined by the
non-linearity in the fractal dimensions spectrum $D(\alpha)$, that gives the fractal dimension of the set.

The main difficulty in validating the model is the lack of a dynamical interpretation
of $D(\alpha)$. An alternative formulation of the model can be made in terms
of probability density functions and the statistical behavior of an ensemble of realizations
of turbulent velocity, rather than its single realization \cite{private}. We will propose that the geometrization of
turbulence may lead to an understanding of $D(\alpha)$ in the language of a single realization of velocity.

\section{Gravity and the Geometrization of Turbulence}

\subsection{Gravity and Holography}

According to the holographic principle, the underlying microscopic degrees of freedom of gravity
in a region of bulk space of volume $V$ are encoded
on a boundary $A$ of the region \cite{'tHooft:1993gx, Susskind:1994vu}.
The origin of the holographic principle is rooted in the physics of black holes, where there is a unique convergence of general relativity, thermodynamics, and quantum field theory. Classical general relativity implies the mechanics of black holes are analogous to the laws of thermodynamics, with the black hole assigned an entropy proportional to its horizon area \cite{Bekenstein:1973ur}. Hawking's discovery shortly thereafter, that black holes radiate like a blackbody and have a real temperature  \cite{Hawking:1974sw}, confirmed that this analogy was in fact an identity, and fixed the black hole entropy to be
\beq S_{BH} = \frac{A}{4 \ell_p^2} \ , \label{bhentropy} \eeq
where $\ell_p$ is the Planck length $(\hbar G c^{-3})^{1/2}$.

In systems with gravity, as more and more matter is packed into a fixed volume, collapse into a black hole is inevitable. Therefore, the Bekenstein-Hawking entropy is the maximal entropy that can be placed in a given volume, meaning that the number of microscopic degrees of freedom is proportional to the area of the region.
This is in sharp contrast to ordinary particle field theories without gravity, where the number of microscopic degrees of freedom is always proportional to the volume of the region. The holographic principle conjecture is rather general and depends neither on the details of microscopic degrees of freedom nor on the emergent bulk geometry.

A particular concrete realization of the holographic principle is provided by the anti-de Sitter (AdS)/conformal field theory (CFT) correspondence \cite{Maldacena:1997re} (for reviews see \cite{reviews}). An AdS spacetime is a negatively curved, hyperbolic spacetime that is solution to the Einstein field equations with a negative cosmological constant. A CFT is a quantum field theory invariant under the conformal group, a symmetry group including scale invariance $x'^\mu = \lambda x^\mu$ and other symmetries, in addition to the usual space-time translations, rotations, and Lorentz boots.  The correspondence relates gravity in $(d+1)$-dimensional AdS spacetimes to certain CFT's living on the $d$-dimensional asymptotic boundary of the spacetime. The microscopic degrees of freedom on the boundary $A$ are identified in terms of a local quantum gauge field theory. Note that in general we expect the microscopic degrees of freedom to be described by a non-local theory. The additional radial coordinate of AdS is interpreted as an emergent dimension corresponding to a coarse graining variable of the microscopic theory, or sometimes simply referred to as an energy scale. Thus, for instance, the size of the object, e.g. instanton in the boundary gauge field theory, is determined by its location in the energy scale direction.

Relativistic hydrodynamics provides a universal description of the large scale dynamics of a CFT.
The AdS/CFT correspondence suggests that the large scale dynamics of gravity in turn provides a dual description of
the CFT hydrodynamics. Indeed, as we will describe below, the four-dimensional CFT hydrodynamics equations are the same as
the equations describing the evolution of large scale perturbations of the five-dimensional black hole \cite{Bhattacharyya:2008jc}.
The limit of non-relativistic macroscopic motions of CFT hydrodynamics
leads to the non-relativistic incompressible NS equations for the ideal and dissipative hydrodynamics of the CFT, respectively \cite{Fouxon:2008tb, Bhattacharyya:2008kq}. Since we can obtain the NS equations in the non-relativistic limit of CFT hydrodynamics, the AdS/CFT correspondence implies that
the NS equations have a dual gravitational description, which can be obtained by taking the non-relativistic limit of the geometry dual to the relativistic CFT hydrodynamics.

\subsection{Black Hole Horizon Dynamics}

We start by considering the Einstein equations with negative cosmological constant $\Lambda$
\beq R_{AB} - \frac{1}{2} g_{AB} R + \Lambda g_{AB} = 8\pi T_{AB}, \label{Einstein} \eeq
$A,B = 0, ...,(d+1)$. Here we use units where the Newton constant, Boltzmann constant, and speed of light are set to unity. $R_{AB}$ and $R$ are the Ricci tensor and scalar, formed from contractions of the Riemann tensor $R_{ABCD}$, which characterizes the {\it intrinsic curvature} of a manifold. Intrinsic curvature can be defined by imagining the parallel transport of a tangent vector around a closed path. If the manifold is curved, the vector's final direction will be different from its initial one; the Riemann tensor measures the amount of deviation. $T_{AB}$ is the stress tensor of a matter source. Therefore the Einstein equations describe the curvature of a spacetime given a matter source.

Here we specialize to five-dimensions ($d=3$) and we will always work with vacuum solutions where $T_{AB} =0$. It is also convenient to choose units where
$\Lambda = -6$. Then (\ref{Einstein}) reduces to
\begin{equation}
R_{AB} + 4 g_{AB} = 0, ~~ R = -20.
\end{equation}
These equations have a black hole solution called a black brane, which can be described by the metric line element
\begin{equation}
ds^2= -r^2 f dt^2+ 2 dt dr + r^2 \sum^{3}_{i=1} dx^i dx_i. \label{equilsolutionrest}
\end{equation}
Here $f=1 - \frac{\pi^4 T^4}{r^4}$ and $x^i = (x,y,z)$. The black hole event horizon is located at $r=\pi T$. Note, that the metric above is invariant under translations in the $x^i$ coordinates. As a result the black brane horizon has the topology of a plane, unlike for example a Schwarzschild black hole, whose horizon is spherical. This can be seen explicitly by evaluating the line element at the horizon. The geometry of the horizon surface is just that of a three-dimensional flat plane.

The black brane solution can be thought of as an equilibrium thermal state at rest at a constant Hawking temperature $T$. It is also possible to make a uniform non-relativistic Galilean translation in the $x^i$ directions, so that the black brane is moving at uniform velocity $v^i$ with respect some rest frame. In this case the metric reads
\begin{eqnarray}
ds^2= -r^2 f dt^2+ 2 dt dr + r^2 \sum^{3}_{i=1} dx^i dx_i  \nonumber \\
-\frac{2 \pi^4 T^4}{r^2} v_i dx^i dt - 2 v_i dx^i dr. \label{equilsolution}
\end{eqnarray}
The AdS/CFT correspondence implies that this is a thermal state of a dual CFT. As for any thermal state, one can study perturbations of this system away from equilibrium. A CFT by definition has no intrinsic scale. Therefore, the only length scale associated with a thermal CFT is the temperature itself, which by dimensional analysis implies that the correlation length of the system must scale as $l_{cor} \sim T^{-1}$. We consider perturbations of characteristic scale $L \gg l_{cor}$, so that the Knudsen number is small and a hydrodynamical description is valid.

To see how the hydrodynamics equations emerge here, it is instructive to consider their derivation in a seemingly different context, from kinetic theory using the Boltzmann equation. A equilibrium Maxwell distribution characterized by constant temperature and velocity is a solution to the Boltzmann equation.  To describe perturbations, one proceeds by the method of variation of the constants, i.e. assuming the velocity and temperature of the distribution are functions of space and time. The new distribution is no longer a solution, but one can introduce corrections and solve for them order by order as a series in Knudsen number. The series solutions can only be constructed if the velocity and temperature satisfy certain constraints. These turn out to be precisely the NS equations.

In the present case, the metric (\ref{equilsolution}) and the Einstein equations are analogous to the Maxwell distribution and Boltzmann equation, respectively. Therefore, we make the ansatz that $v^i(t,x^i)$, $T(t,x^i)$ and look for a
solution of the Einstein equation of the form
\begin{equation}
g_{AB}=(g_0)_{AB}+\delta g_{AB} \ ,
\end{equation}
where
\begin{eqnarray}
(g_0)_{AB}dy^{A}dy^{B}= -r^2 f(t,x^i) dt^2+ 2 dt dr + r^2 \sum^{3}_{i=1} dx^i dx_i  \nonumber \\
-\frac{2 \pi^4 T(t,x^i)^4}{r^2} v_i(t,x^i) dx^i dt - 2 v_i(t,x^i) dx^i dr \ , \label{zerothorder} \end{eqnarray}
and $y = (t, x^{i}, r)$.  The term $\delta g_{AB}$ symbolizes the lowest order corrections in Knudsen number and velocity.  As a bookkeeping device one can introduce the parameter $\varepsilon$ and the scaling
\beq v^i \sim \partial_i \sim \varepsilon, ~ \partial_t \sim \varepsilon^2 \ . \label{epsilonscaling} \eeq
With this choice of scaling, (\ref{zerothorder}) has terms of $O(1)$ and $O(\varepsilon)$. The correction $\delta g_{AB}$ should have pieces of $O(\varepsilon^2)$ and $O(\varepsilon^3)$. We assume the temperature has the expansion
\begin{equation}
T = T_0(1+P(t,x^i)) \ ,
\end{equation}
where the scalar quantity $P$ is of $O(\varepsilon^2)$ and is ultimately identified as a non-relativistic pressure.

The solution of the Einstein equations (up to $O(\varepsilon^4)$) was constructed in \cite{Bhattacharyya:2008kq} by taking a non-relativistic scaling limit of the earlier, fully relativistic results in \cite{Bhattacharyya:2008jc}. As in the Boltzmann equation, the condition of constructibility of the series solution requires that the velocity and pressure satisfy the incompressible NS equations, but with a particular kinematic viscosity $\nu = (4\pi T_0)^{-1}$. The NS equations are four out of the five Einstein equations that constrain ``initial data" on surfaces of constant $r$. The rest of the equations describe the ``dynamics" of the geometry in the radial direction.

Given a solution to the incompressible NS equations, i.e. a viscous flow of the boundary CFT,  there is dual five-dimensional gravitational solution
(see e.g Eq. (\ref{equilsolution})) . The existence of the regular black brane horizon is crucial on the gravitational side. The classical boundary condition associated with the horizon - that fields may fall into the black hole, but cannot emerge from it - breaks time reversal symmetry and explains how the Einstein equations can describe dissipative effects. The increase of entropy in the fluid is manifested geometrically in the dual description by the increase in the cross-sectional area of the black hole horizon \cite{Bhattacharyya:2008xc}. This is consistent with the Bekenstein-Hawking black hole entropy (\ref{bhentropy}), which obeys a second law of thermodynamics.

The way the black brane horizon geometry encodes the boundary fluid dynamics is reminiscent of the {\it Membrane Paradigm} in classical general relativity
\cite{Damour, Price, membrane}, according to which any black hole has a fictitious viscous fluid living on its horizon. Although the real fluid, whose dynamics we wish to study, is typically thought of as being on the hologram at the boundary of the spacetime, it is natural to ask to what extent it can be identified under the duality map with the membrane paradigm fluid on the horizon.

Using a formalism developed by Damour \cite{Damour}, the dynamics of the event horizon of a black brane in the Membrane Paradigm has been analyzed
and shown to be described by the incompressible Navier-Stokes equations  \cite{Eling:2009pb}. Hence, there is a direct mapping between the fluid variables of
an incompressible flow and the geometrical variables characterizing a horizon. More generally, the fluid variables of an incompressible flow in $(d+1)$ spacetime dimensions ($d \geq 2$) with $\nu= (4\pi T_0)^{-1}$ can be mapped into the geometrical variables of a $(d+1)$ dimensional horizon surface.

The analysis performed in  \cite{Eling:2009pb} is quite general, and holds for any non-singular null hypersurface as long as a large scale hydrodynamics limit exists.  It is important to stress, however, that the analysis does not apply to the familiar Schwarzschild black hole.  In the Schwarzschild solution there is only one scale, the mass $M$, or equivalently the inverse Hawking temperature $T^{-1}$. For a compact object, the characteristic size $L$ of its perturbations is limited by its radius $r_0$, which here must $\sim M$.  Thus, the Knudsen number is always of order unity, an expansion in derivatives is not valid, and there is no hydrodynamic limit. On the other hand,
the analysis applies to the Rindler horizon perceived by an accelerated observer in flat Minkowksi spacetime (which has no intrinsic length scale) even though there is no known specific holographic duality like AdS/CFT in this case.

We now review the details of \cite{Eling:2009pb}, focusing on the black brane example. To keep the discussion  general, we consider a $(d+2)$-dimensional bulk space-time $M$ with coordinates $X^A, A=0,...,d+1$ and a Lorentzian metric $g_{AB}$. Let $H$ be a $(d+1)$-dimensional null hypersurface in $M$. A null hypersurface is special because its normal vector ${\bf n}$ satisfies,
\begin{equation}
 {\bf n} \cdot {\bf n} =
g_{AB} n^A n^B = 0 \ . \label{normal}
\end{equation}
This condition is equivalent to the statement that the normal vector is also a tangent vector in null hypersurface.

We define the hypersurface in the bulk space-time by
$x^{d+1} \equiv r = const$, and denote the other coordinates
as $x^{\mu} = (t, x^i), i=1,...,d$. The coordinate $t$ parameterizes a slicing of space-time by spatial hypersurfaces and
$x^i$ are coordinates on sections of the horizon with constant $t$. Generally in this coordinate system
\begin{equation}
n^r =0,~~~~n^t=1,~~~~ n^i = v^i \ . \label{normalform}
\end{equation}
In the case of black brane metric above (\ref{equilsolution}), the event horizon in equilibrium is located in the bulk space-time
at $r=\pi T_0$, where $T_0$ corresponds to the constant Hawking temperature. The normal vector to the horizon is (\ref{normalform}). We can interpret the normal vector here as a $(d+1)$-velocity, with the components $v^i$ as the non-relativistic flow velocity with respect to a rest frame defined by the sections of constant $t$.  As before, we work in the slow motion limit where $v^i$ is a small perturbation and use the scaling given in (\ref{epsilonscaling}).

The first important quantity, which describes the geometry of the null hypersurface is the {\it induced metric}. This is the pullback of  $g_{AB}$ to the horizon, which in practice means the its components are projected into the horizon surface and evaluated at $r= const.$. In the general coordinate system,
\beq ds^2_H = h_{ij} (dx^i - v^i dt) (dx^j - v^j dt) \ ,
\eeq
where $h_{ij}$ is the metric on sections $S_t$ of the horizon $H$ at
constant $t$. For the equilibrium black brane, at zeroth order in $\varepsilon$
\beq h^{(0)}_{ij} = (\pi T_0)^2 \delta_{ij} \ .
\eeq
The details of the subleading terms, which are of order $O(\varepsilon^2)$ and higher, will not be needed for the analysis.

One needs one additional geometrical object in order to have a complete description of how the null hypersurface is embedded in the bulk space-time. This second fundamental form of the horizon hypersurface is called the {\it extrinsic curvature} $K^\mu_\nu$. This object is different from the intrinsic curvature and characterizes the bending of a hypersurface in a bulk spacetime. For example, a cylinder is not intrinsically curved; parallel transport of tangent vectors around closed paths on the surface leads to no deviation. However, it is clearly curved when seen as a surface embedded in a higher dimensional space. In general this notion is described by the surface components (here represented by $\mu$) of the derivative of the normal vector in the embedding space (see Figure 2)

\begin{figure}[htb]
\begin{center}
\epsfig{file=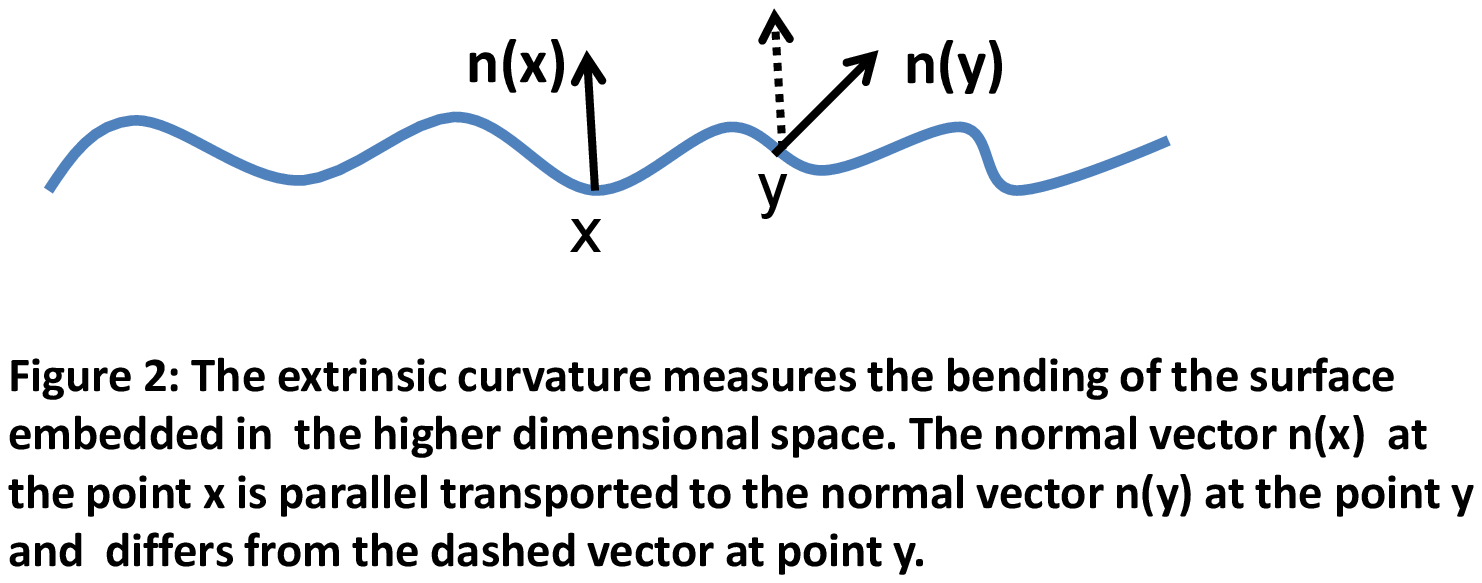,height=11cm}
\end{center}
\vskip -6cm
\end{figure}

Using $\bf n$ and $e^A_{i}$ as a tangent basis, the components
of the horizon's extrinsic curvature are
\begin{equation}
K^{n}_{n} = \kappa(x),~~~~~~
K^{n}_{i} = \Omega_i,~~~~~~
K^{i}_j = \sigma^i_j + \theta \delta^i_j/d  \ .
\end{equation}
$\kappa(x)$ is  called the ``surface gravity",  because in a static, equilibrium spacetime it is the force that must be exerted at infinity in order to hold the object stationary at the horizon.
It can be parameterized as
\beq \kappa(x) = 2 \pi
T_0(1 + P(x)) \ ,  \label{kappa} \eeq
where $P(x)$ scales as $\varepsilon^2$. We will identify $P(x)$ as the
fluid pressure.
The horizon location is $r_H = \pi T_0 (1+P(x))$ (see Figure 3).
\begin{figure}[htb]
\begin{center}
\epsfig{file=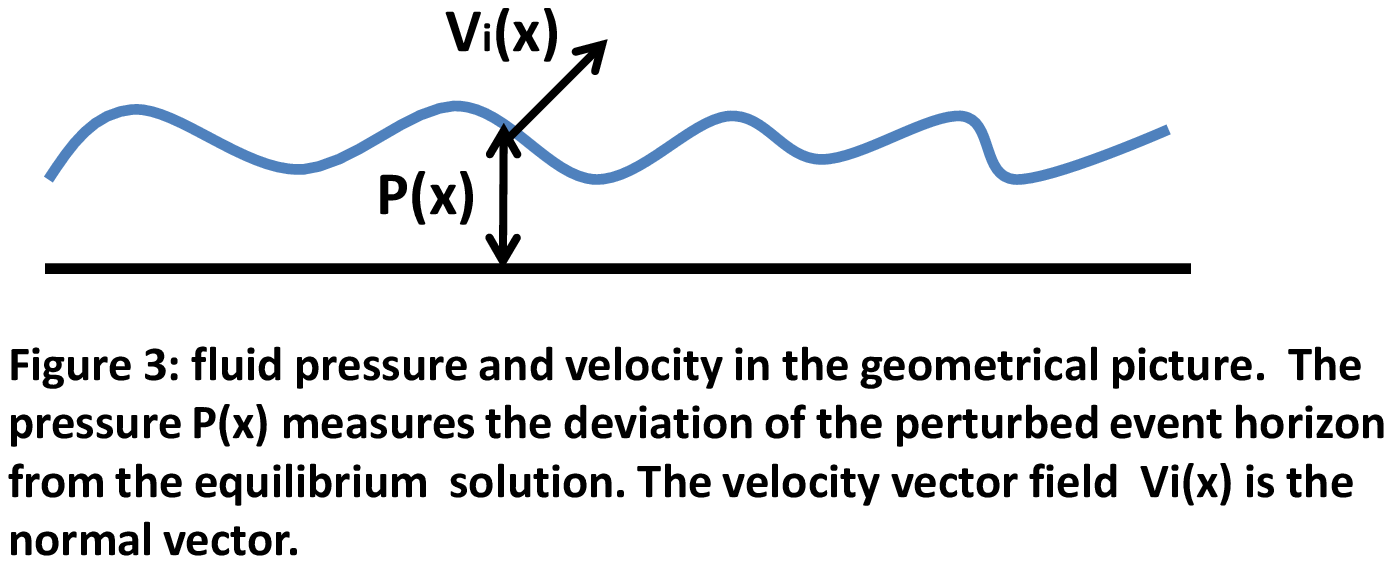,height=11cm}
\end{center}
\vskip -6cm
\end{figure}

$\Omega_i$ can be interpreted as a ``momentum" for the horizon.
For the black brane, at leading order we have
\beq \Omega_i = 2 \pi T_0 v_i \ .  \eeq

The extrinsic curvature components $K^i_j$  can be split into trace and trace-free parts, which are the two most important geometrical objects. These are the {\it horizon expansion} $\theta$,  and the {\it horizon shear} $\sigma_{ij}$, respectively. Given a bundle of light rays, the expansion is the fractional rate of change of the cross-sectional area, while the shear tensor describes deformations to the bundle that preserve the cross-sectional area (see Figure 4).  For the black brane we get to leading order the $O(\varepsilon^2)$ expressions
\bea \theta &=& \partial_i v_i \\
\sigma_{ij} &=&  \half (\pi T_0)^2 (\partial_i v_j+\partial_j v_i -  
2\partial_k v_k \delta_{ij}/d ) \ .
\eea
We can already see from these expressions a close link between the horizon geometrical expansion and shear and their fluid counterparts. The horizon expansion is the fluid expansion, while the horizon shear is equal to the fluid shear, up to the proportionality constant $(\pi T_0)^2 $.
\begin{figure}[htb]
\begin{center}
\epsfig{file=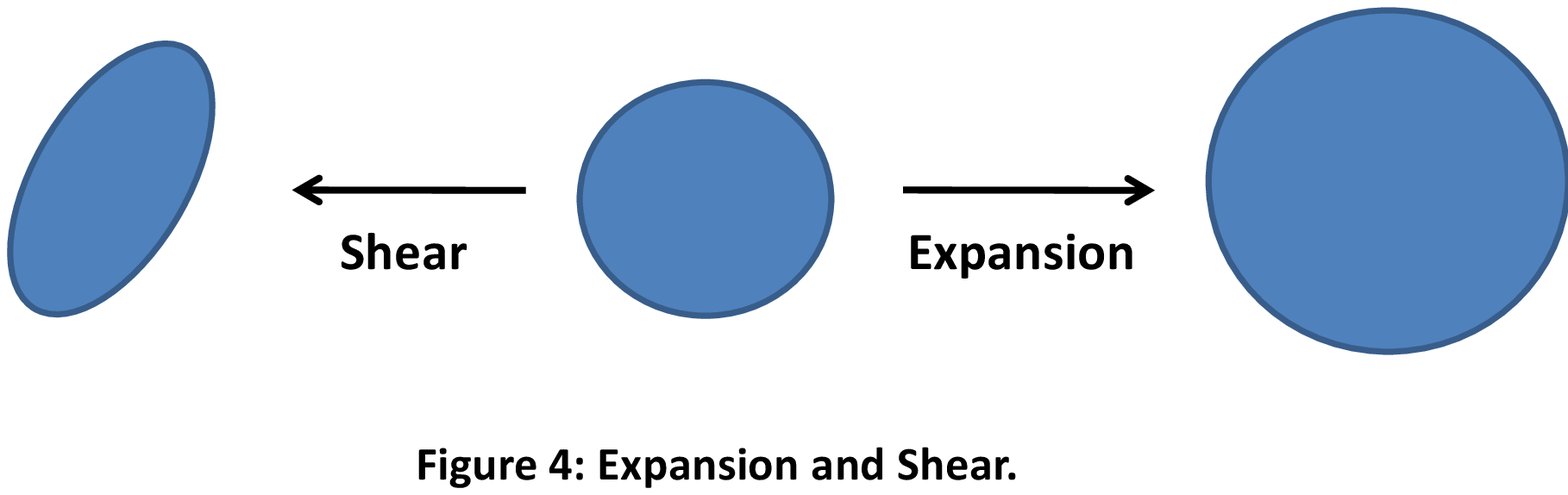,height=10cm}
\end{center}
\vskip -7cm
\end{figure}
The dynamics of the horizon geometry perturbations
are governed by the Einstein equations. As we saw above, the black brane is a
particular solution to the Einstein equations with negative cosmological
constant, here in a general spacetime dimension, $R_{AB} + (d+1) g_{AB} = 0 $.
We want to study the set of $(d+1)$ Einstein equations projected into the null horizon surface. Consider first the contraction with $n^A n^B$.  There is no contribution from the cosmological constant
term proportional to the metric due to (\ref{normal}). The resulting equation can be expressed solely in terms of the horizon geometrical variables as the null focusing equation
\beq - \partial_t \theta - v^i \partial_i \theta + \kappa(x) \theta - \theta^2/d -
\sigma_{ij} \sigma^{ij}  = 0 \ .  \label{focusing} \eeq
This equation describes the time evolution of the expansion, essentially how a pencil of light rays converges in the gravitational field. Plugging our previous results for the expansion, shear, and surface gravity of the black brane, we found that at leading order the focusing equation is equivalent to the incompressibility condition
\beq \partial_i v_i = 0 \ .
\eeq

Now consider the contraction of the Einstein equations with
$n^A e^B_i$. This is a more complicated equation for the evolution of the momentum $\Omega_i$.
Again it can be expressed in terms of the geometrical variables and, as before, the cosmological constant term does not
contribute. At the lowest orders in $\varepsilon$, it sufficient to express the equation simply in terms of partial derivatives of the horizon geometrical variables,
\beq (\partial_t + \theta)\Omega_i  +
v^j \partial_j \Omega_i + \Omega_j \partial_i v^j = -\partial_i
\kappa(x) + \partial_j \sigma^j_i -
\frac{1}{d}
\partial_i \theta. \label{horizonNS}\eeq
Plugging in the results for $\Omega_i$,  $\theta$,  $\sigma_{ij}$ and $\kappa$, we found that the leading order terms are
at order $\varepsilon^3$ and give precisely NS equation (\ref{NS})
with a kinematic viscosity $\nu = (4\pi T_0)^{-1}$,  but without a force term. A forcing term can be added by considering a black brane solution to the Einstein equations in the presence of an in-falling matter field, say for instance, a scalar field.

From the fluid NS equations one can always derive an energy balance law relating the change of the kinetic energy to minus the rate of viscous dissipation,
\beq \int \partial_t  v^2/2 ~ d^{d} x = - \nu \int \partial_i v_j \partial^i v^j d^d x \ . \eeq
In the geometric picture we can use the focusing equation to relate the rate of change in kinetic energy in turn to an increase in the horizon area,
\beq \partial_t \left(A/A_0\right) =
- \int \partial_t  v^2/2 ~ d^{d} x \ ,  \label{areadiss}\eeq
where $A_0$ is the zeroth order area density $(\pi T_0)^d$. Thus, as we expect from second law of black hole thermodynamics, as
the kinetic energy of the fluid on the boundary decreases in time
due to viscous dissipation, the horizon area grows. It is interesting to note that when the fluid is in inertial range, where viscous effects are small, our relation implies that in the gravitational/geometrical picture the horizon area is nearly constant.

The mapping we have found between dynamics of a surface and the NS
equation is surprising. It is reminiscent of the connection between the Burgers and the KPZ equations. The $(1+1)$-dimensional
Burgers equation \cite{Burgers} provides a simplified model for turbulence, while the KPZ equation \cite{Kardar:1986xt} describes a local growth of an interface using a height function $h(x,t)$. The height gradient obeys the Burgers equation
\begin{equation}
\partial_t v + v \partial_x v - \nu \partial_{xx} v  = f,~~~~~~~v=\partial_x h \ .
\end{equation}
Our result shows that real turbulence may also be seen as resulting from a physically natural surface dynamics.

Finally, we should note that a similar analysis can be performed for relativistic fluids relating the
hydrodynamics equations to horizon dynamics \cite{Eling:2009sj,Bhattacharyya:2008xc}.
This can provide a geometrical language to study universality aspects of relativistic turbulence statistics \cite{FO2}.

\section{Outlook}

We described the emergence of a new language to characterize the motion of fluids - the language of
space-time geometry that is used in the theory of general relativity. We argued that motions of Newtonian fluids are contained
within a sub-class of the solutions to  Einstein equations without sources. In general, a benefit of a new way of looking at a problem is in
suggesting new methods for solving it; methods that would otherwise be hard to find or use. We can thus ask,
what aspects of the problem of turbulence could be expected to be addressed more efficiently using the new formulation.

The geometrical formulation may improve the rationale for the phenomenological multi-fractal model that was described before. The original formulation of the model was in terms of a single realization of
the velocity, and it referred to the hypothetical singularities of the realizations. A firmer formulation of the
model is achieved by formulating it as a hypothesis on the statistics of the realizations, rather than on a single realization \cite{Frisch,private}. While the latter formulation is more sound and robust, the question on the dynamical meaning of the model remains open. 
This issue is important
for trying to connect the model with the governing NS equations, and for providing a theoretical basis for the model. In particular, it is
not clear what are the sets in space that are multi-fractal. The general relativistic formulation suggests a natural answer to the above question.
The hypothesis that correlation functions of velocity differences, such as $S_n(r)$, obey anomalous scalings in the inertial
range means in the General Relativistic picture that space-averaged equal-time correlation functions of differences of normals to the horizon surface
obey anomalous scalings in the inertial range. This strongly suggests that the horizon surface itself is multi-fractal for turbulence in the inertial range, thus
providing a natural geometrical structure for the study of the hypothesis of multi-fractality.

 A second problem in the study of turbulence, where the geometrical formulation can help, is the finite-time singularity structure of the NS equations. Here we address the problem in three space dimensions. It is still not known whether the Cauchy problem is well-posed for the NS equations, i.e. whether a unique solution exists for all times given smooth initial conditions. In particular, it is
an open question whether the evolution prescribed by the equations produces singularities in velocity gradients within a finite time.
In fact, 
these key questions are on the list of the Clay Institute mathematical problems, mentioned before. The understanding of the counterpart of this
problem for the Einstein equations - the issue of naked singularities - is much better understood. This is thanks to powerful geometrical methods that have been developed in the theory of general relativity
\cite{MTW}. The disparity between our understanding of singularities in the theories of turbulence and general relativity, and the fact that the former is essentially contained within the latter, gives hope that the General Relativistic formulation of the problem of turbulence may lead to a major advancement in the study of singularities of the NS equations.

Will the General Relativistic formulation produce actual advancement in the study of turbulence?
The formulation is still less than three years old, while decades of studies by traditional means resulted with a little progress in the study of the NS equations.
Thus, we hope that such a radical change in  the way of thinking about turbulence, as has been considered here, may turn out to be very
useful. Evidently, only extensive studies will tell us.

\section*{Acknowledgements}

The work of Y.O. is supported in part by the Israeli
Science Foundation center of excellence, by the Deutsch-Israelische
Projektkooperation (DIP), by the US-Israel Binational Science
Foundation (BSF), and by the German-Israeli Foundation (GIF).

\end{document}